\documentclass[aps,12pt,final,oneside,onecolumn,nobibnotes,nofootinbib,%
superscriptaddress,centertags,floatfix,secnumarabic,notitlepage]{revtex4}%
\usepackage{graphicx}
\usepackage{amssymb,amsmath}
\usepackage{bm}
\usepackage{slashed}
\usepackage{subfigure}
\usepackage{multirow}
\RequirePackage[english]{babel}
\usepackage{cleveref}
\usepackage[normalem]{ulem}
\usepackage{xcolor}
\usepackage{amsmath}
\usepackage{amsfonts}
\usepackage{setspace}
\usepackage{amssymb}
\newcommand{\be}{\begin{equation}}
\newcommand{\ee}{\end{equation}}
\newcommand{\bdis}{\begin{displaymath}}
\newcommand{\edis}{\end{displaymath}}
\newcommand{\bga}{\begin{equation}\begin{gathered}}
\newcommand{\ega}{\end{gathered}\end{equation}}
\newcommand{\mathsym}[1]{{}}
\newcommand{\unicode}[1]{{}}

\begin{document}

\selectlanguage{english}
\normalsize

\title{Higgs boson decay to paired $B_c$: relativistic and one-loop corrections.}

\author{\firstname{I.~N.}~\surname{Belov}}
\email{ilia.belov@cern.ch}
\affiliation{SINP MSU, Moscow, Russia}
\affiliation{Physics department of MSU, Moscow, Russia}

\author{\firstname{A.~V.}~\surname{Berezhnoy}}
\email{Alexander.Berezhnoy@cern.ch}
\affiliation{SINP MSU, Moscow, Russia}

\author{\firstname{A.~E.}~\surname{Dorokhov}}
\email{dorokhov@theor.jinr.ru}
\affiliation{Joint Institute of Nuclear Research, BLTP, Moscow region, Dubna, Russia}

\author{\firstname{A.~K.}~\surname{Likhoded}}
\email{Anatolii.Likhoded@ihep.ru}
\affiliation{NRC ''Kurchatov Institute'' IHEP, Protvino, Russia}

\author{\firstname{A.~P.}~\surname{Martynenko}}
\email{a.p.martynenko@samsu.ru}
\affiliation{Samara University, Samara, Russia}

\author{\firstname{F.~A.}~\surname{Martynenko}}
\email{f.a.martynenko@gmail.com}
\affiliation{Samara University, Samara, Russia}

\begin{abstract}
\small

The exclusive decays of the Higgs boson  to  $B_c B_c$ and $B_c^\ast B_c^\ast$ pairs  are studied. 
The hard parts of the decay amplitudes are estimated within the perturbative Standard Model up to 
one-loop corrections.   The soft fusion of heavy quarks to the quarkonium is described 
in framework of the relativistic quark model.
\end{abstract}

\maketitle

\normalsize
\section{Introduction}

One of the main goals of the research programs of the CMS and ATLAS experiments at the LHC is to study the properties of the Higgs boson~\cite{atlas,cms}. The decays into two $W$ bosons, two $Z$ bosons and two photons play a key role in the Higgs boson investigation. Among other decays, the decays into two quarkonia, including the decays into the pair of pseudoscalar mesons $B_c B_c$ or 
vector mesons $B_c^\ast B_c^\ast$, may be of particular interest 
(note, that  the Higgs boson decay process $H\to B_c^\ast+B_c$ is forbidden due to the conservation 
law of angular momentum).
Such processes allow to study the coupling constants of the Higgs boson with heavy quarks, as well as to test the theory of bound states of heavy quarks produced in decays. In this work, we calculate both the relativistic corrections connected with the relative motion of heavy quarks and the QCD 
one-loop corrections to the Higgs boson decay width. 
As is well known from the studies of various authors, starting with the production of $J/\psi$ and
$\eta_c$ mesons, the corrections of both these types significantly contribute to the production cross 
section of a pair of heavy quarkonia. 

A bound state $(\bar b c)$ with open beauty and charm has a special place among the heavy quarkoniums  since its decay mechanism differs significantly from the decay mechanism of charmonium or bottomonium. That is why we believe that  the process $H\to B_c+B_c$ will attract the attention of experimenters. 

Our approach to the calculation of the observed Higgs boson decay widths leading to the pair 
production of $B_c$ mesons is based on the methods of relativistic quark model (RQM) and the perturbative Standard Model \cite{apm5,apm3}.
This approach allows a systematic account for relativistic effects throughout
the construction of relativistic amplitudes of pair production of mesons, relativistic production 
cross sections, and in the description of bound states of quarks themselves through the use of the corresponding quark interaction potential.

As it is known from the previous study~\cite{Berezhnoy:2016etd}, the paired $B_c$ production is essentially affected by one loop QCD corrections. Thus we 
take them into account within the same method, as was applied in \cite{Berezhnoy:2016etd}
using modern computer methods for calculating the Feynman interaction amplitudes.

One of the first works devoted to the pair production of quarkonia in Higgs boson decays was done 
in the nonrelativistic approximation in \cite{keung}. 
The production of single quarkonia in the H decay was investigated in \cite{vysotsky,bodwin} 
with the account
of relativistic corrections and one-loop corrections.
In the work \cite{luchinsky}, various channels of the Higgs boson decay into pairs of heavy 
quarkonia were studied, including $H\to ZZ$ and $H\to WW$.
The single $B_c$ meson production rate in Higgs boson decays was calculated within the  
nonrelativistic QCD framework in \cite{qiao}.
The first experimental searches for decays of the Higgs boson into a pair of $J/\Psi$ 
and $\Upsilon$ mesons  were performed in \cite{cms1}.

While the quarkonia with a hidden flavour have been studied experimentally well enough, 
the experimental data on the bound states of heavy quark and antiquark with different flavours are rather poor \cite{bll2019}. In fact, such states, $B_c$ mesons, 
are known for the $1S$ and $2S$ states  only. Therefore, the study of various mechanisms for the production of $B_c$ mesons is of obvious interest, which is connected with the study of their properties.

\section{General RQM formalism}

\begin{figure}[htbp]
\centering
\includegraphics[scale=0.6]{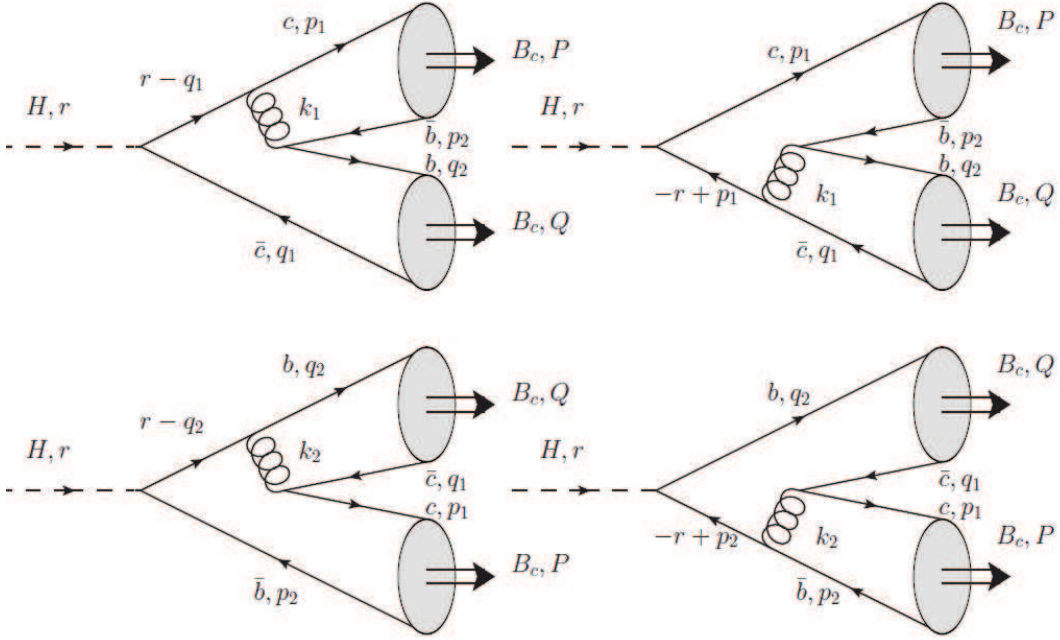}
\caption{The pair $B_c$-meson production amplitudes in Higgs boson decay. $B_{c}$
denotes the $B_c$-meson states with spin 0 and 1. Dashed line shows the Higgs boson
and wavy line corresponds to the gluon.}
\label{fig1}
\end{figure}

Four production amplitudes of the $B_c$ meson pair in leading order of the QCD coupling constant $\alpha_s$
are presented in Fig.~\ref{fig1}.
We investigate the production channel of a pair of $B_c$ mesons connected with the initial production 
of a pair of heavy quarks $b$ or $c$ in the Higgs boson decay.
There are two stages of $B_c$ meson production process. At the first stage, which is described
by the perturbative Standard Model, the Higgs boson transforms into a heavy quark-antiquark pair. Then
the heavy quark or antiquark emits a virtual gluon which produces another heavy quark-antiquark pair.
At the second stage, heavy quarks and antiquarks combine with some probability into bound states.

Four-momenta of heavy quarks and antiquarks can be expressed in terms of relative and total 
four momenta as follows:
\begin{equation}
\label{eq:pq}
p_1=\eta_{1}P+p,~p_2=\eta_{2}P-p,~(p\cdot P)=0,~\eta_{i}=\frac{M_{B_c}^2\pm m_1^2\mp m_2^2}{2(M_{B_c})^2},
\end{equation}
\begin{displaymath}
q_1=\rho_{1}Q+q,~q_2=\rho_{2}Q-q,~(q\cdot Q)=0,~\rho_{i}=\frac{M_{B_c}^2\pm m_1^2\mp m_2^2}{2(M_{B_c})^2},
\end{displaymath}
where $M_{B_c}$ is the mass of pseudoscalar or vector $B_c^+$ ($B_c^{\ast +}$) meson
consisting of $\bar b$-antiquark and $c$-quark.
$m_{1,2}$ are the masses of $c$ and $b$ quarks. 
$P(Q)$ are the total four-momenta of mesons $B_c^+$ and $B_c^{\ast -}$, relative quark four-momenta
$p=L_P(0,{\bf p})$ and
$q=L_P(0,{\bf q})$ are obtained from the rest frame four-momenta $(0,{\bf p})$ and $(0,{\bf q})$ by the
Lorentz transformation to the system moving with the momenta $P$ and $Q$.
The index $i=1,2$ corresponds to plus and minus signs in \eqref{eq:pq}.
Heavy quarks $c$, $b$ and antiquarks $\bar c$, $ \bar b$ in the intermediate state are outside the mass shell:
$p_{1,2}^2=\eta_{i}^2P^2-{\bf p}^2=\eta_{i}^2M_{B_c}^2-{\bf p}^2\not= m_{1,2}^2$,
so that $p_1^2-m_1^2=p_2^2-m_2^2$.

Let consider the production amplitude of pseudoscalar and vector $B_c$ mesons. Initially it can be written as a
convolution of perturbative production amplitude of free quarks and antiquarks and the quasipotential wave
functions of $B_c$ mesons moving with four-momenta P and Q. Using then the transformation 
law of the bound state wave functions from the rest frame to the
moving one with four-momenta $P$ and $Q$ we can present the meson production amplitude in the form
\cite{apm3,apm1,apm2}:
\begin{equation}
\label{eq:amp}
{\cal M}(p_-,p_+,P,Q)=-iM(\sqrt{2}G_F)^{1/2}\frac{2\pi}{3}M_{B_c}
\int\frac{d{\bf p}}{(2\pi)^3}
\int\frac{d{\bf q}}{(2\pi)^3}
\times
\end{equation}
\begin{displaymath}
\times Sp\left\{\Psi^{\cal P,V}_{B_c}(p,P)\Gamma_1^{\nu}(p,q,P,Q)\Psi^{\cal P,V}_{B_c}(q,Q)
\gamma_\nu+
\Psi^{\cal P,V}_{B_c}(-p,P)\Gamma_2^{\nu}(p,q,P,Q)\Psi^{\cal P,V}_{B_c }(-q,Q)
\gamma_\nu\right\},
\end{displaymath}
where
a superscript ${\cal P}$ indicates a pseudoscalar $B_c$ meson, a superscript ${\cal V}$ indicates a vector
$B_c$ meson, $G_F$ is the Fermi constant, $M=m_1+m_2$. $\Gamma_{1,2}$ are the vertex functions defined below.
The permutation of subscripts $b$ and $c$ in the wave functions indicates corresponding permutation
in the projection operators (see below Eqs.\eqref{eq:amp1}-\eqref{eq:amp2}.
The method for producing the amplitudes in the form \eqref{eq:amp} is described in detail in our previous studies
\cite{apm3,apm4,apm5}.
The transition of free quark-antiquark pair to meson bound states is described in our approach by specific wave functions.
Relativistic wave functions of pseudoscalar and vector $B_c$ mesons accounting for the transformation from the
rest frame to the moving one with four momenta $P$, and $Q$ are
\begin{eqnarray}
\label{eq:amp1}
\Psi^{\cal P}_{B_c}(p,P)&=&\frac{\Psi^0_{B_c}({\bf p})}{
\sqrt{\frac{\epsilon_1(p)}{m_1}\frac{(\epsilon_1(p)+m_1)}{2m_1}
\frac{\epsilon_2(p)}{m_2}\frac{(\epsilon_2(p)+m_2)}{2m_2}}}
\left[\frac{\hat v_1-1}{2}+\hat
v_1\frac{{\bf p}^2}{2m_2(\epsilon_2(p)+ m_2)}-\frac{\hat{p}}{2m_2}\right]\cr
&&\times\gamma_5(1+\hat v_1) \left[\frac{\hat
v_1+1}{2}+\hat v_1\frac{{\bf p}^2}{2m_1(\epsilon_1(p)+
m_1)}+\frac{\hat{p}}{2m_1}\right],
\end{eqnarray}
\begin{eqnarray}
\label{eq:amp2}
\Psi^{\cal V}_{B^\ast_c}(q,Q)&=&\frac{\Psi^0_{B^\ast_c}({\bf q})}
{\sqrt{\frac{\epsilon_1(q)}{m_1}\frac{(\epsilon_1(q)+m_1)}{2m_1}
\frac{\epsilon_2(q)}{m_2}\frac{(\epsilon_2(q)+m_2)}{2m_2}}}
\left[\frac{\hat v_2-1}{2}+\hat v_2\frac{{\bf q}^2}{2m_1(\epsilon_1(q)+
m_1)}+\frac{\hat{q}}{2m_1}\right]\cr &&\times\hat{\varepsilon}_{\cal
V}(Q,S_z)(1+\hat v_2) \left[\frac{\hat v_2+1}{2}+\hat
v_2\frac{{\bf q}^2}{2m_2(\epsilon_2(q)+ m_2)}-\frac{\hat{q}}{2m_2}\right],
\end{eqnarray}
where the symbol hat denotes convolution of four-vector with the Dirac gamma matrices,
$v_1=P/M_{B_c}$, $v_2=Q/M_{B_c}$;
$\varepsilon_{\cal V}(Q,S_z)$ is the polarization vector of the $B^{\ast-}_c(1^-)$ meson,
relativistic quark energies $\epsilon_{1,2}(p)=\sqrt{{\bf p}^2+m_{1,2}^2}$.
Relativistic functions~\eqref{eq:amp1}-\eqref{eq:amp2}
and the vertex production functions $\Gamma_{1,2}$
do not contain the $\delta ({\bf p}^2-\eta_{i}^2M_{B_c}^2+m_{1,2}^2)$ which corresponds 
to the transition on the mass shell.
In~\eqref{eq:amp1} and \eqref{eq:amp2} we have complicated factor including the bound state wave 
function in the rest frame.
Therefore instead of the substitutions $M_{B_c}=\epsilon_1({\bf p})+\epsilon_2({\bf p})$ and
$M_{B^\ast_c}=\epsilon_1({\bf q})+\epsilon_2({\bf q})$ in the production amplitude we carry out the
integration over the quark relative momenta ${\bf p}$ and ${\bf q}$.
The color part of the meson wave function in the amplitude~\eqref{eq:amp} is taken as $\delta_{ij}/\sqrt{3}$
(color indexes $i, j, k=1, 2, 3$).
Relativistic wave functions in~\eqref{eq:amp1} and \eqref{eq:amp2} are equal to the product of wave functions 
in the rest frame
$\Psi^0_{B_c}({\bf p})$ and spin projection operators that are
accurate at all orders in $|{\bf p}|/m$. An expression of spin projector in different
form for $(c\bar c)$ system was obtained in \cite{bodwin2002} where spin projectors are
written in terms of heavy quark momenta $p_{1,2}$ lying on the mass shell. Our derivation of relations~\eqref{eq:amp1} and \eqref{eq:amp2} accounts for the transformation law of the bound state wave functions from the rest frame to the
moving one with four momenta $P$ and $Q$. This transformation law was discussed in the Bethe-Salpeter approach in \cite{brodsky} and in quasipotential method in \cite{faustov}.

We have omitted here intermediate expressions, leading to the equations~\eqref{eq:amp}-\eqref{eq:amp2}
because they were discussed in detail in our previous papers.
In the Bethe-Salpeter approach the initial production amplitude has as a form of convolution of the truncated
amplitude with two Bethe-Salpeter (BS) $B_c$ meson wave functions.
The presence of the $\delta (p\cdot P)$ function in this case
allows us to make the integration over relative energy $p^0$. In the rest frame of a bound state the condition
$p^0=0$ allows to eliminate the relative energy
from the BS wave function. The BS wave function satisfies a two-body bound state equation
which is very complicated and has no known solution. A way to deal with this problem
is to find a soluble lowest-order equation containing main physical properties
of the exact equation and develop a perturbation theory. For this purpose we continue
to work in three-dimensional quasipotential approach. In this framework the double
$B_c$ meson production amplitude~\eqref{eq:amp} can be written initially as a product of the production
vertex function $\Gamma_{1,2}$ projected onto the positive energy states by means of the Dirac
bispinors (free quark wave functions) and a bound state quasipotential wave functions
describing $B_c$ mesons in the reference frames moving with four momenta $P,Q$.
Further transformations include the known transformation law of the bound state wave
functions to the rest frame \cite{apm3,apm4}. 
In the spin projectors we have
${\bf p}^2\not=\eta_{i}^2M^2-m_{1,2}^2$ just the same as in the vertex production functions
$\Gamma_{1,2}$.
We can consider \eqref{eq:amp1}-\eqref{eq:amp2} as a transition form factors for
heavy quark-antiquark pair from free state to bound state. When transforming the amplitude ${\cal M}$
we introduce the projection operators $\hat\Pi^{\cal P,V}$ onto the states of $(Q_1\bar Q_2)$ in the $B_c$ meson
with total spin 0 and 1 as follows:
\begin{equation}
\label{eq:uu}
\hat\Pi^{\cal P,V}=[v_2(0)\bar u_1(0)]_{S=0,1}=\gamma_5(\hat\varepsilon^\ast)\frac{1+\gamma^0}{2\sqrt{2}}.
\end{equation}

At leading order in $\alpha_s$ the vertex functions $\Gamma_{1,2}^{\nu}(p,P;q,Q)$ can be written as
($\Gamma_2^{\nu}(p,P;q,Q)$ can be obtained from $\Gamma_1^{\nu}(p,P;q,Q)$ by means of the replacement
$p_1\leftrightarrow p_2$, $q_1\leftrightarrow q_2$, $\alpha_b\to\alpha_c$)
\begin{equation}
\label{eq:g1}
\Gamma_1^{\nu}(p,P;q,Q)= r_1\alpha_b\left[\gamma_\mu\frac{(\hat r-\hat
q_1+m_1)}{(r-q_1)^2-m_1^2+i\epsilon} +
\frac{(\hat p_1-\hat r+m_1)}{(p_1-r)^2-m_1^2+i\epsilon}
\gamma_\mu\right]D^{\mu\nu}(k_1),
\end{equation}
\begin{equation}
\label{eq:g2}
\Gamma_2^{\nu}(p,P;q,Q)=r_2\alpha_c\left[\gamma_\mu\frac{(\hat r-\hat q_2+m_2)}{(r-q_2)^2-m_2^2+i\epsilon}+
\frac{(\hat p_2-\hat r+m_2)}{(p_2-r)^2-m_2^2+i\epsilon}
\gamma_\mu\right]D^{\mu\nu}(k_2),
\end{equation}
where $r^2=M_H^2=(P+Q)^2=2M_{B_c}^2+2PQ$,
the gluon four-momenta are $k_1=p_1+q_1$, $k_2=p_2+q_2$, $\alpha_{c,b}=\alpha_s\left(m_{1,2}^2M_H^2/(m_1+m_2)^2\Lambda^2\right)$.
Relative momenta $p$, $q$ of heavy quarks enter in the gluon propagators $D_{\mu\nu}(k_{1,2})$
and quark propagators as well as in relativistic wave functions~\eqref{eq:amp1} and \eqref{eq:amp2}.
Accounting for the small ratio of relative quark momenta $p$ and $q$ to the mass $M_H$, we use an expansion of
inverse denominators of quark and gluon propagators as follows:
\begin{equation}
\label{den1}
\frac{1}{(p_1+q_1)^2}=\frac{1}{\eta_1 \rho_1 M_H^2},~~~
\frac{1}{(p_2+q_2)^2}=\frac{1}{\eta_2 \rho_2 M_H^2},
\end{equation}
\begin{equation}
\label{den2}
\frac{1}{(r-q_1)^2-m_1^2}=\frac{1}{\rho_2 M_H^2},~~~
\frac{1}{(-r-p_1)^2-m_1^2}=\frac{1}{\eta_2 M_H^2},
\end{equation}
\begin{equation}
\label{den3}
\frac{1}{(r-p_2)^2-m_1^2}=\frac{1}{\eta_1 M_H^2},~~~
\frac{1}{(-r-q_2)^2-m_1^2}=\frac{1}{\rho_1 M_H^2}.
\end{equation}

Using expansions \eqref{den1}-\eqref{den3} and wave functions \eqref{eq:amp1}-\eqref{eq:amp2}
in the amplitude~\eqref{eq:amp} we hold the second-order correction for small ratios
$|{\bf p}|/m_{1,2}$, $|{\bf q}|/m_{1,2}$, $|{\bf p}|/M_H$, $|{\bf q}|/M_H$ relative to the leading order result.
As we take relativistic factors in the denominator of the amplitudes \eqref{eq:amp1} and \eqref{eq:amp2} unchanged,
the momentum integrals are convergent. Calculating the trace in obtained expression
in the package FORM \cite{form}, we find relativistic amplitudes of the $B_c$ meson pairs production
in the form:
\begin{equation}
\label{eq:amp11}
{\cal M_{PP}}=\frac{32\pi}{3M_H^4}(\sqrt{2}G_F)^{1/2}M_{B_c}M
\left[\frac{\alpha_{b}r_1}{\eta_2^3}F_{1P}+\frac{\alpha_{c}r_2}{\eta_1^3}F_{2P}\right]
|\tilde\Psi_P(0)|^2,
\end{equation}
\begin{equation}
\label{eq:amp22}
{\cal M_{VV}}=\frac{32\pi}{3M_H^4}(\sqrt{2}G_F)^{1/2}M_{B_c}M \varepsilon_1^{\lambda}\varepsilon_2^{\sigma}
\left[\frac{\alpha_{b}r_1}{\eta_2^3}F_{1V}^{\lambda\sigma}+\frac{\alpha_{c}r_2}
{\eta_1^3}F_{2V}^{\lambda\sigma}\right]|\tilde\Psi_V(0)|^2,
\end{equation}
where $\varepsilon_{{\cal V}}$ is the polarization vector of spin 1 $B_c$ meson.
The decay widths of the Higgs boson into a pair of pseudoscalar and vector $B_c$ mesons are determined
by the following expressions:
\begin{equation}
\label{gapp}
\Gamma_{PP}=\frac{512\sqrt{2}\pi G_F M^2 |\tilde\Psi_P(0)|^4 \sqrt{\frac{r_3^2}{4}-1}}{9M_H^5 r_3^5}
\left[\frac{\alpha_{b}r_1}{\eta_2^3}F_{1P}+\frac{\alpha_{c}r_2}{\eta_1^3}F_{2P}\right]^2,
\end{equation}
\begin{equation}
\label{gapp1}
F_{1P}=-r_1-\eta_1+\frac{3}{2}r_3^2-\frac{1}{2}r_1 r_3^2-\frac{1}{2}\eta_1 r_3^2+\omega_{01}
(-12r_2+2r_2r_3^2)+
\end{equation}
\begin{displaymath}
\omega_{10}(2r_1+r_1r_3^2)+\omega_{10}\omega_{01}(6r_2-2r_1-\frac{3}{2}r_3^2-
r_3^2r_2-r_3^2r_1),
\end{displaymath}
\begin{equation}
\label{gavv}
\Gamma_{VV}=\frac{512\sqrt{2}\pi G_F M^2 |\tilde\Psi_V(0)|^4 \sqrt{\frac{r_3^2}{4}-1}}{9M_H^5 r_3^5}
\sum_{\lambda,\sigma}
\vert\varepsilon_1^\lambda\varepsilon_2^\sigma
\left[\frac{\alpha_{b}r_1}{\eta_2^3}F^{\lambda\sigma}_{1V}+\frac{\alpha_{c}r_2}{\eta_1^3}
F^{\lambda\sigma}_{2V}\right]\vert^2,
\end{equation}
\begin{equation}
\label{gavv1}
F^{\alpha\beta}_{1V}=g_1 v_1^\alpha v_2^\beta+g_2g^{\alpha\beta},~~~
F^{\alpha\beta}_{2V}=\tilde g_1 v_1^\alpha v_2^\beta+\tilde g_2g^{\alpha\beta},
\end{equation}
\begin{displaymath}
g_1=-1+\frac{1}{9}\omega_{10}\omega_{01},~~~g_2=-r_1-\eta_1+\frac{1}{2}r_3^2-\frac{4}{3}r_2\omega_{01}+
2r_1\omega_{10}+\omega_{10}\omega_{01}(\frac{2}{3}r_2-\frac{2}{9}-\frac{1}{18}r_3^2).
\end{displaymath}
In the nonrelativistic limit the Higgs boson decay rates acquire the form:
\begin{multline}
\Gamma_{PP}^{nr}=\frac{512\sqrt{2}\pi G_F M^2 |\tilde\Psi_P(0)|^4 \sqrt{\frac{r_3^2}{4}-1}}{9M_H^5 r_3^5} \times \\
\times \Biggl[\frac{\alpha_{b}r_1}{\eta_2^3}\Bigl( \frac{3}{2}r_3^2-2r_1-r_1 r_3^2 \Bigr)+\frac{\alpha_{c}r_2}{\eta_1^3}
\Bigl( \frac{3}{2}r_3^2-2r_2-r_2 r_3^2 \Bigr) \Biggr]^2,
\label{eq:LO-PP}
\end{multline}
\begin{multline}
\Gamma_{VV}^{nr}=\frac{512\sqrt{2}\pi G_F M^2 |\tilde\Psi_V(0)|^4 \sqrt{\frac{r_3^2}{4}-1}}{9M_H^5 r_3^5}
\frac{1}{4r_1^6 r_2^6}\times \\
\times\Biggl[\alpha_{b}^2 r_1^8 \Bigl(4 r_1^2 \left(r_3^4-4r_3^2+12\right)-4 
r_1 \left(r_3^2+2\right) r_3^2+3 r_3^4\Bigr)+\\
+2 \alpha_{b} \alpha_{c} r_1^4 r_2^4 \Bigl(4 r_1 r_2 \left(r_3^4-
4r_3^2+12\right)-2 r_1 \left(r_3^2+2\right)r_3^2-2 r_2 \left(r_3^2+
2\right)r_3^2+3 r_3^4\Bigr)+ \\
+\alpha_{c}^2 r_2^8 \Bigl(4 r_2^2 \left(r_3^4-4
r_3^2+12\right)-4 r_2 \left(r_3^2+2\right) r_3^2+3 r_3^4\Bigr)\Biggr]
\label{eq:LO-VV}
\end{multline}
where the parameter $r_3=\frac{M_H}{M_{B_c}}$.

The functions $F_{i P}$,  $F_{i V}$ entering in \eqref{eq:amp11}-\eqref{eq:amp22}
can be written initially as series in specific relativistic factors $C_{ij}=[(m_1-\epsilon_1(p))/(m_1+\epsilon_1(p))]^i
[(m_2-\epsilon_2(q))/(m_2+\epsilon_2(q))]^j$ with $i+j\leq 2$ connected with
the relative momenta $p$ and $q$ of heavy quarks.
In final form the functions $F_{i P}$,  $F_{i V}$ and the production cross sections 
contain relativistic parameters $\omega^{P,V}_{nk}$ which are determined by 
the momentum integrals $I_{nk}$ and calculated in the quark model:
\begin{equation}
\label{eq:intnk}
I_{nk}^{P,V}=\int_0^\infty q^2R^{P,V}_{B_c}(q)\sqrt{\frac{(\epsilon_1(q)+m_1)(\epsilon_2(q)+m_2)}
{2\epsilon_1(q)\cdot 2\epsilon_2(q)}}
\left(\frac{m_1-\epsilon_1(q)}{m_1+\epsilon_1(q)}\right)^n
\left(\frac{m_2-\epsilon_2(q)}{m_2+\epsilon_2(q)}\right)^k dq,
\end{equation}
\begin{equation}
\label{eq:parameter}
\omega^{P,V}_{10}=\frac{I^{P,V}_{10}}{I^{P,V}_{00}},~\omega^{P,V}_{01}=\frac{I^{P,V}_{01}}{I^{P,V}_{00}},~
\omega^{P,V}_{20}=\frac{I^{P,V}_{20}}{I^{P,V}_{00}},
\omega^{P,V}_{02}=\frac{I^{P,V}_{02}}{I^{P,V}_{00}},~\omega^{P,V}_{11}=\frac{I^{P,V}_{11}}{I^{P,V}_{00}},
\end{equation}
\begin{equation}
\tilde\Psi^0_{B_c}(0)=\int \sqrt{\frac{(\epsilon_1(p)+m_1)(\epsilon_2(p)+m_2)}
{2\epsilon_1(p)\cdot 2\epsilon_2(p)}}\Psi^0_{B_c}({\bf p})\frac{d{\bf p}}{(2\pi)^3}.
\end{equation}

Another source of relativistic corrections is related with the Hamiltonian of the heavy quark bound states
which allows to calculate 
the bound state wave functions of pseudoscalar and vector $B_c$ mesons.
The exact form of the bound state wave function $\Psi^0_{B_c}({\bf q})$
is important to obtain more reliable predictions for the decay widths. 
In nonrelativistic approximation the pair $B_c$ meson production cross sections
contain fourth power of nonrelativistic wave function at the origin. The value of the cross sections is very sensitive 
to small changes of $\Psi^0_{B_c}$. In nonrelativistic QCD there exists corresponding problem 
of determining the magnitude of the
color-singlet matrix elements \cite{bbl}. To account for relativistic corrections to the meson wave functions
we describe the dynamics of heavy quarks by the QCD generalization of the standard Breit Hamiltonian in the center-of-mass reference frame 
\cite{repko1,pot1,capstick,godfrey,glko,godfrey1,rqm1}:
\begin{equation}
\label{eq:breit}
H=H_0+\Delta U_1+\Delta U_2,~~~H_0=\sqrt{{\bf
p}^2+m_1^2}+\sqrt{{\bf p}^2+m_2^2}-\frac{4\tilde\alpha_s}{3r}+(Ar+B),
\end{equation}
\begin{equation}
\label{eq:breit1}
\Delta U_1(r)=-\frac{\alpha_s^2}{3\pi r}\left[2\beta_0\ln(\mu
r)+a_1+2\gamma_E\beta_0
\right],~~a_1=\frac{31}{3}-\frac{10}{9}n_f,~~\beta_0=11-\frac{2}{3}n_f,
\end{equation}
\begin{equation}
\label{eq:breit2}
\Delta U_2(r)=-\frac{2\alpha_s}{3m_1m_2r}\left[{\bf p}^2+\frac{{\bf
r}({\bf r}{\bf p}){\bf p}}{r^2}\right]+\frac{2\pi
\alpha_s}{3}\left(\frac{1}{m_1^2}+\frac{1}{m_2^2}\right)\delta({\bf r})+
\frac{4\alpha_s}{3r^3}\left(\frac{1}{2m_1^2}+\frac{1}{m_1m_2}\right)({\bf S}_1{\bf L})+
\end{equation}
\begin{displaymath}
+\frac{4\alpha_s}{3r^3}\left(\frac{1}{2m_2^2}+\frac{1}{m_1m_2}\right)({\bf S}_2{\bf L})
+\frac{32\pi\alpha_s}{9m_1m_2}({\bf S}_1{\bf S}_2)\delta({\bf r})+
\frac{4\alpha_s}{m_1m_2r^3}\left[\frac{({\bf S}_1{\bf r})({\bf S}_2{\bf r})}{r^2}-
\frac{1}{3}({\bf S}_1{\bf S}_2)\right]-
\end{displaymath}
\begin{displaymath}
-\frac{\alpha_s^2(m_1+m_2)}{m_1m_2r^2}\left[1-\frac{4m_1m_2}{9(m_1+m_2)^2}\right],
\end{displaymath}
where ${\bf L}=[{\bf r}\times{\bf p}]$, ${\bf S}_1$, ${\bf S}_2$ are spins of heavy quarks,
$n_f$ is the number of flavors, $\gamma_E\approx 0.577216$ is
the Euler constant. To improve an agreement of theoretical hyperfine splittings in $(\bar bc)$ mesons
with experimental data and other calculations in quark models we add to the standard Breit potential \eqref{eq:breit2}
the spin confining potential obtained in \cite{repko1,repko2,gupta,gupta1}:
\begin{equation}
\Delta V^{hfs}_{conf}(r)=
f_V\frac{A}{8r}\left\{\frac{1}{m_1^2}+\frac{1}{m_2^2}+\frac{16}{3m_1m_2}({\bf S}_1{\bf S}_2)+
\frac{4}{3m_1m_2}\left[3({\bf S}_1 {\bf r}) ({\bf S}_2 {\bf r})-({\bf S}_1 {\bf S}_2)\right]\right\},
\end{equation}
where we take the parameter $f_V=0.9$. For the dependence of the
QCD coupling constant $\tilde\alpha_s(\mu^2)$ on the renormalization point
$\mu^2$ in the pure Coulomb term in~\eqref{eq:breit} we use the three-loop result \cite{kniehl1997}
\begin{equation}
\label{26}
\tilde\alpha_s(\mu^2)=\frac{4\pi}{\beta_0L}-\frac{16\pi b_1\ln L}{(\beta_0 L)^2}+\frac{64\pi}{(\beta_0L)^3}
\left[b_1^2(\ln^2 L-\ln L-1)+b_2\right], \quad L=\ln(\mu^2/\Lambda^2).
\end{equation}
In other terms of the Hamiltonians~\eqref{eq:breit1} and \eqref{eq:breit2} we use
the leading order approximation for $\alpha_s$. The typical momentum transfer scale in a
quarkonium is of order of double reduced mass, so we set the renormalization scale $\mu=2m_1m_2/(m_1+m_2)$ and
$\Lambda=0.168$ GeV, which gives $\alpha_s=0.265$ for $(\bar bc)$ meson.
The coefficients $b_i$ are written explicitly in \cite{kniehl1997}.
The parameters of the linear potential $A=0.18$ GeV$^2$ and $B=-0.16$ GeV have established values in quark models.

\begin{table}[h]
\caption{Numerical values of relativistic parameters~\eqref{eq:parameter}
and decay widths of Higgs boson \eqref{gapp}, \eqref{gavv} with the account of 
relativistic corrections.}
\bigskip
\label{tb1}
\begin{ruledtabular}
\begin{tabular}{|c|c|c|c|c|c|c|c|c|c|}
$B_c$ &$n^{2S+1}L_J$ &$M_{B_c}$, &$\Psi^0_{B_c}(0)$, & $\omega^{P,V}_{10}$ &$\omega^{P,V}_{01}$
 &  $\omega^{P,V}_{20}$ &$\omega^{P,V}_{02}$ &  $\omega^{P,V}_{11}$ & $\Gamma_{nr}$, in GeV \\
meson  &     &   GeV  &   GeV$^{3/2}$  &   &  &   &  &  &   $\Gamma_{rel}$, in GeV  \\    \hline
$B_c$&$1^1S_0$ & 6.275 & 0.250 & -0.0489 & -0.0060 & 0.0049& 0.0001  & 0.0006 &$0.56\cdot 10^{-12}$\\
&  &  &   &   &   &   &   &   &   $0.29\cdot 10^{-12}$  \\  \hline
$B^\ast_c$  & $1^3S_1$  & 6.317 & 0.211 & -0.0540  & -0.0066   & 0.0053  &  0.0001   
& 0.0007 &$0.56\cdot 10^{-12}$  \\
&  &  &   &   &   &   &   &   &   $0.15\cdot 10^{-12}$  \\  \hline
\end{tabular}
\end{ruledtabular}
\end{table}

The numerical values of the relativistic parameters entering the cross 
sections~\eqref{gapp}, and \eqref{gavv}
are obtained by the numerical solution of the Schr\"odinger equation \cite{LS}.
They are collected in Table~\ref{tb1} in which we present also the results of
nonrelativistic and relativistic calculation of Higgs boson decay widths.

\section{One loop corrections}

Estimating NLO corrections we calculate LO widths using the workflow, which differs from one 
applied within RQM.  Following both workflows we have obtained the same expressions 
\eqref{eq:LO-PP} and \eqref{eq:LO-VV} for LO widths; this served as a cross-check of our 
calculations. Another check is the explicitly obtained zero for the prohibited process of 
$B_c^*B_c$ production at both LO and NLO levels.

\begin{figure}[ht]
\centering
\includegraphics[width=0.98\linewidth]{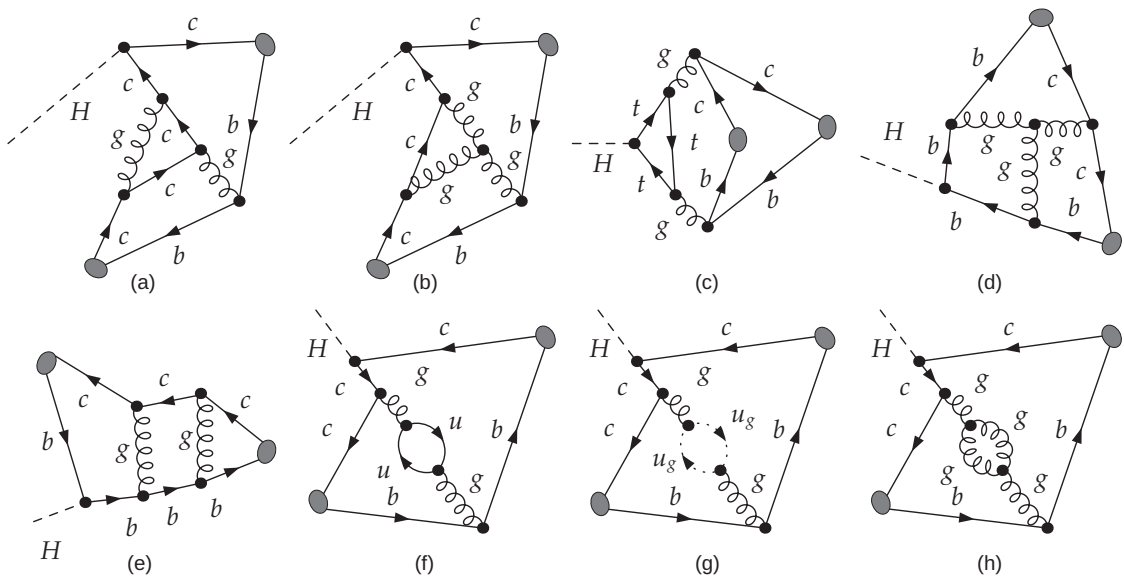}
\caption{Typical one loop diagrams for the paired $B_c$-meson production in  the Higgs boson decay.}
\label{fig:diagsNLO}
\end{figure}

Up to the NLO accuracy ${\cal O}(\alpha_s^3)$ the matrix element squared is expressed as follows:

\begin{equation}\label{eq:amp-NLO}
|{\cal M}|^2 = |{\cal M}_{LO}|^2  + 2Re\left({\cal M}_{LO}{\cal M}_{NLO}^*\right).
\end{equation}

The Higgs boson decay to paired $B_c$ is described by a set of 86 diagrams at next-to-leading order. The typical diagrams are shown in Fig.~\ref{fig:diagsNLO}.  

The computation strategy is based on the following toolchain in Wolfram Mathematica: \texttt{FeynArts}~\cite{Hahn:2000kx} $\to$ \texttt{FeynCalc}~\cite{Shtabovenko:2020gxv} (\texttt{FeynCalcFormLink}~\cite{Feng:2012tk}, \texttt{TIDL}) $\to$ \texttt{Apart}~\cite{Feng:2012iq} $\to$ \texttt{FIRE}~\cite{Smirnov:2008iw} $\to$ \texttt{X}-package~\cite{Patel:2016fam}. 
The amplitudes generated with the \texttt{FeynArts} package are further processed with \texttt{FeynCalc} package, which provides algebraic calculations with Dirac and color matrices, including the evaluation of traces. The Passarino-Veltman reduction is carried out using the  \texttt{TIDL} library implemented in \texttt{FeynCalc}. The \texttt{Apart} function does the extra simplification of the integrals. The \texttt{FIRE} package provides the complete reduction of  the  integrals obtained in the previous stages  to master integrals, using the IBP reduction strategy mostly based on the Laporta algorithm~\cite{Laporta:2001dd}.  The master integrals are then evaluated by substitution of their analytical expressions with the help of \texttt{X}-package.

The conventional dimensional regularization (CDR) scheme with $D$-dimensional momenta  (loop and external) and Dirac matrices was used.  $\gamma^5$ is known to be poorly defined in $D$ dimensions. However $\gamma^5$ matrices are canceled out in the amplitude of $H\to B_c B_c$ decay and are completely absent in the amplitude of $H\to B_c^* B_c^*$ decay. Therefore we do not face the problem of $\gamma^5$ definition estimating the one loop QCD corrections for the discussed processes.  

After the \texttt{FIRE} reduction  only one-, two- and three-point integrals ($\boldsymbol{A}_0$, 
$\boldsymbol{B}_0$, $\boldsymbol{C}_0$) are left in the amplitudes. Some integrals of types $\boldsymbol{A}_0$ and  $\boldsymbol{B}_0$ contribute to the amplitude with the singular coefficient 
$~\frac{1}{D-4}$. In such cases we should keep  the terms of the order of ${\cal O}(\varepsilon)$ 
in the master integral expansion over  $\varepsilon$, because  these terms might contribute 
to the finite part of an amplitude  (see \cite{Berezhnoy:2016etd} for details).

The so-called ``On shell'' scheme is adopted for masses and spinors renormalization and $\overline{MS}$ scheme is adopted for coupling constant renormalization:
\begin{align}\label{mrenorm}
Z_m^{OS} &= 1 - \frac{\alpha_s}{4\pi}C_F C_{\epsilon}\left[\frac{3}{\epsilon_{UV}} + 4\right] + O(\alpha_s^2), \\ \label{wfrenorm}
Z_2^{OS} &= 1 - \frac{\alpha_s}{4\pi}C_F C_{\epsilon}\left[\frac{1}{\epsilon_{UV}} + \frac{2}{\epsilon_{IR}} + 4\right] + O(\alpha_s^2), \\ \label{grenorm}
Z_g^{\overline{MS}} &= 1 - \frac{\beta_0}{2}\frac{\alpha_s}{4\pi}\left[\frac{1}{\epsilon_{UV}} -\gamma_E + \ln(4\pi)\right] + O(\alpha_s^2),
\end{align}
where $C_{\epsilon} = \left(\frac{4\pi\mu^2}{m^2}e^{-\gamma_E}\right)^{\epsilon}$ and  $\gamma_E$ is the Euler constant.

The isolated singularities in the one loop amplitude ${\cal \widetilde M}_{NLO}$ are further 
cancelled with singular parts of ${\cal M}_{CT}$ so that 
${\cal M}_{NLO}= {\cal \widetilde M}_{NLO}+{\cal M}_{CT}$ remains a finite expression 
for the renormalized amplitude, where

\begin{equation}
{\cal M}_{CT} = Z_2^2{\cal M}_{LO} \Biggr|_{\substack{\boldsymbol{ m \to Z_m m} \\
\boldsymbol{g_s \to Z_g g_s}}}.
\end{equation}

Note that the amplitudes of the pair production of $B_c$ mesons with the emission 
of one soft gluon  vanish if we describe $B_c$ mesons in the color singlet model.

\begin{figure}[ht]
\centering
\begin{minipage}[h]{0.48\linewidth}\label{fig:MuDependence}
\centering
\includegraphics[width = 1.05\linewidth]{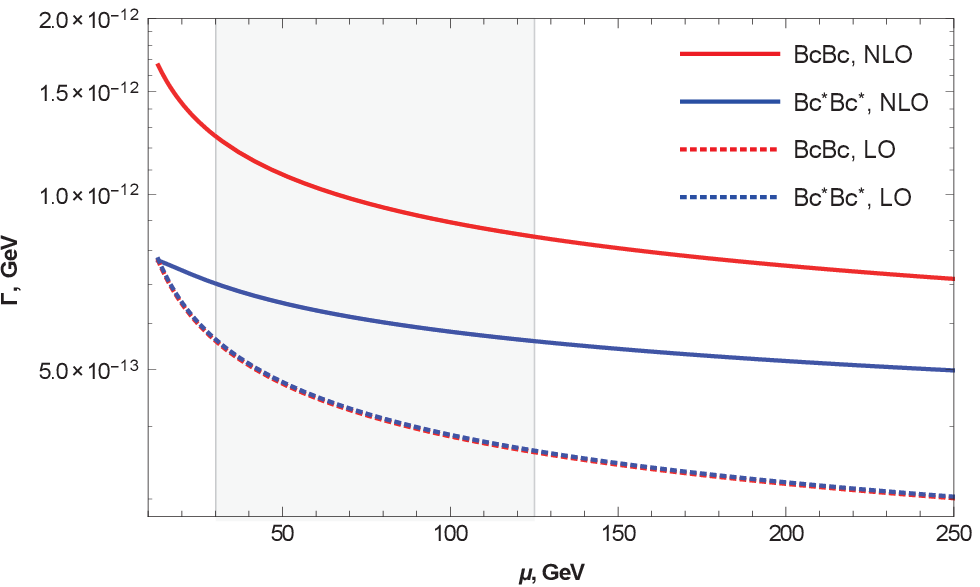}
\end{minipage}
\hfill
\begin{minipage}[h]{0.48\linewidth}\label{fig:RatioMuDependence}
\centering
\includegraphics[width = 1\linewidth]{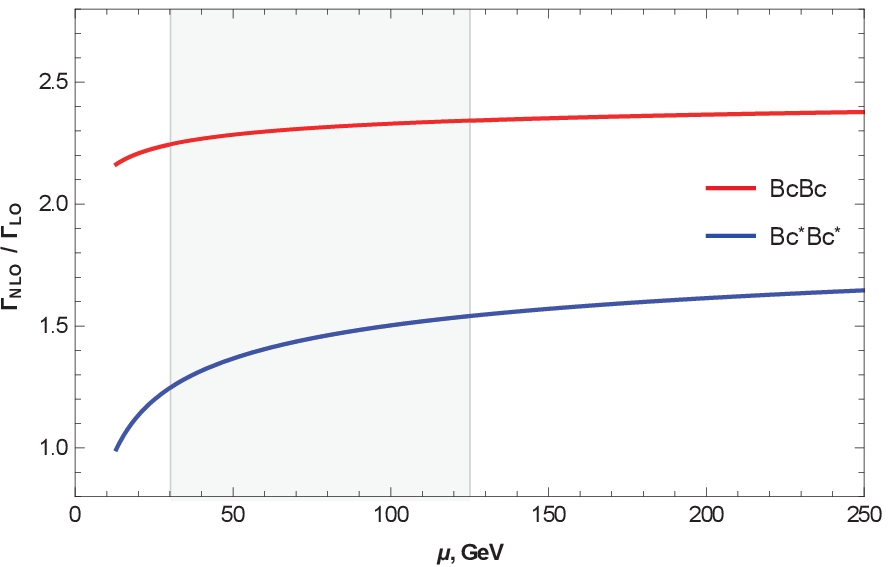}
\end{minipage}
\vfill
\caption{Scale dependence of the decay widths estimated within LO and NLO: 
the absolute values (left) and the  NLO/LO ratio (right). The filled area displays 
the range $m_1 M_H/(m_1+m_2) \le \mu \le M_H$. } 
\label{fig3} 
\end{figure}

\begin{figure}[ht]
\centering
\begin{minipage}[h]{0.48\linewidth}\label{fig:RDependence}
\centering
\includegraphics[width = 1.05\linewidth]{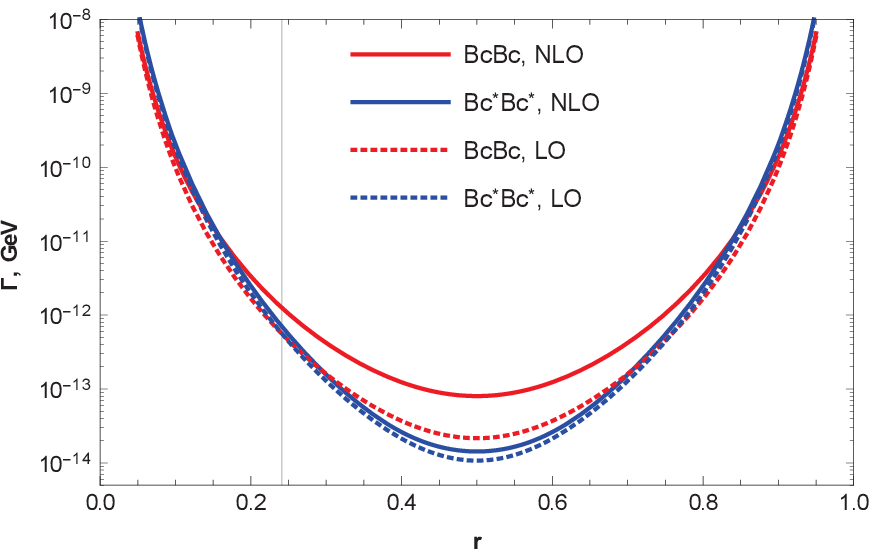}
\end{minipage}
\hfill
\begin{minipage}[h]{0.48\linewidth}\label{fig:RatioRDependence}
\centering
\includegraphics[width = 1\linewidth]{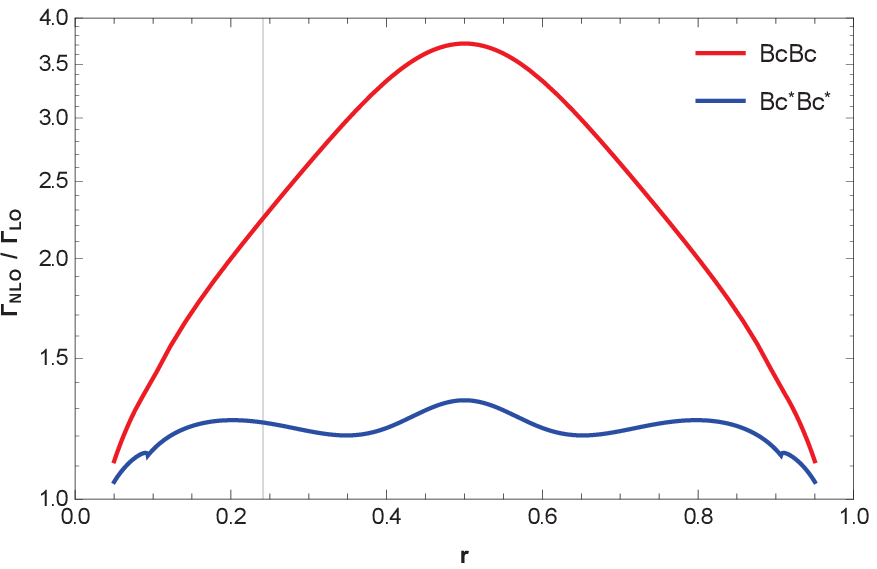}
\end{minipage}
\vfill
\caption{Dependence of the decay widths estimated within LO and NLO on $r=m_1/(m_1+m_2)$:  
absolute values (left) and  NLO/LO ratio (right). The vertical line corresponds 
to $r=1.55/6.43$.}
\label{fig4}
\end{figure}

The calculation results are shown in the Fig.~\ref{fig3} where the dependence on the scale choice is presented for the LO and NLO approaches, and in the Fig.~\ref{fig4}, 
where the dependence on the quark mass values is demonstrated. 
Numerical values of the NLO decay widths at different energy scales are
presented in Table~\ref{tb2}.

As it can be seen in the Fig.~\ref{fig3},  the NLO/LO ratio quite slowly depends on the scale choice  
both for $B_c B_c$ and $B_c^*B_c^*$. Also it is interesting that the one loop corrections essentially change the ratio of the decay widths between  $B_c B_c$ and $B_c^*B_c^*$ pairs: while the LO approach predicts the ratio $\Gamma_{VV}/\Gamma_{PP}\sim 1$, the NLO corrections increase this value to 
$1.5\div 1.9$ depending on the scale choice. 

As it is demonstrated in the Fig.~\ref{fig4}, the predictions can be sensitive 
to the choice of the quark mass ratio $m_1/(m_1+m_2)$. However, we do not focus on these details 
in the current study, as we  think that the problem of mass choice deserves a separate 
consideration (see for example \cite{Kataev:1993be, Kataev:2009ns}, where the problem 
of choice of the  $b$ quark mass was studied in details). 

\begin{figure}[htbp]
\centering
\includegraphics[scale=1.]{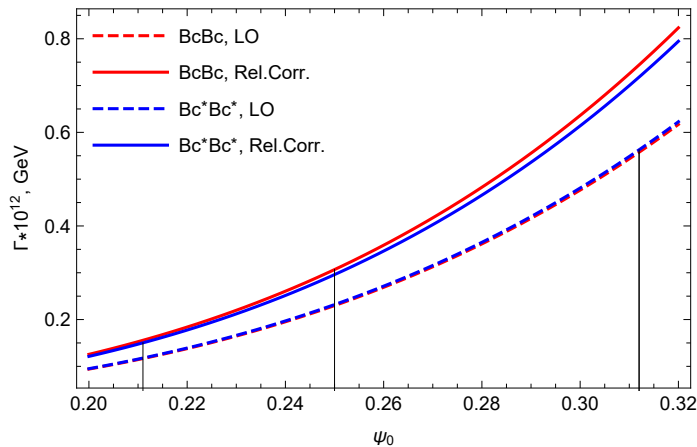}
\caption{The dependence of the $H$ boson decay width on the value of the wave 
function of the $B_c$ meson at zero $\psi_0=\Psi^0_{B_c}(0)$. The vertical lines
correspond to values of $\psi_0$ in nonrelativistic approximation and with the account
of relativistic corrections.}
\label{fig5}
\end{figure}

\begin{table}[ht]
\caption{NLO decay widths at different scales. The renormalization scale $\mu_R$ and the strong coupling scale $Q$ are chosen equal: $\mu=\mu_R=Q$.}
\bigskip
\label{tb2}
\begin{tabular}{|c|c|c|c|c|c|}
\hline
& \multirow{2}{*}{$\Gamma_{LO}$, GeV} & \multicolumn{4}{c|}{$\Gamma_{NLO}$, GeV} \\ \cline{3-6}
 &   & $m_1M_H/(m_1+m_2)$  & $M_H/2$ & $m_2M_H/(m_1+m_2)$  & $M_H$  \\    \hline
$B_c B_c$ & $0.56\cdot 10^{-12}$  & $1.26\cdot 10^{-12}$ & $1.01\cdot 10^{-12}$ & $0.91\cdot 10^{-12}$ &  $0.85\cdot 10^{-12}$  \\  
NLO/LO & ---  &  2.24 & 2.30 & 2.33  & 2.34  \\ \hline
$B^\ast_c B^\ast_c$  & $0.56\cdot 10^{-12}$  & $0.70\cdot 10^{-12}$  & $0.63\cdot 10^{-12}$ & $0.59\cdot 10^{-12}$ & $0.56\cdot 10^{-12}$  \\ 
NLO/LO & ---  & 1.25  & 1.41 & 1.49  & 1.54   \\ \hline\hline\hline
\end{tabular}
\end{table}

\section{Conclusion}

In this work, we investigate the process of pair production of $B_c$ mesons in the decay of the 
Higgs boson. 
The exact amplitudes of the decay of the $H$ boson into a pair of scalar and vector $B_c$ mesons 
are constructed, 
in which the dependence on the relative momenta of heavy quarks (relativistic corrections) is taken into account. 
Then the approximate decay amplitudes are obtained, in which the second-order relativistic 
corrections are 
retained. The widths of the decay of the $H$ boson are also calculated both in the nonrelativistic approximation 
and with an account of the second-order relativistic corrections. Along with relativistic corrections in decay amplitudes, we also take into account second-order relativistic corrections when calculating the wave functions 
of $B_c$ mesons in the framework of the relativistic quark model. The obtained analytical expressions for the decay 
widths are used to carry out numerical estimates. As in the solution of various previous problems on the pair 
production of bound quark states in \cite{apm1,apm2,apm3,apm4}, our calculations in this work show 
that relativistic effects are very important 
for finding reliable values of the Higgs boson decay widths. Taking into account all relativistic corrections 
leads to a decrease in the decay widths of $\Gamma_{PP}$ and $\Gamma_{VV} $ by several times. 
The main factor leading to this decrease is the value of the wave function of quark bound states at zero. 
Our calculations show that relativistic corrections reduce this value by 30 percent, and since 
the decay widths \eqref{gapp} and \eqref{gavv} include the fourth power $|\psi^0_{B_c}(0)|$, 
the decrease in the decay widths themselves turns out to be very significant (see the results 
in Table~\ref{tb1}).
The change in the $H$ boson decay width depending on $|\psi^0_{B_c}(0)|$ is shown 
in a separate Fig.~\ref{fig5}
due to the importance of this nonperturbative factor.
In this work, a purely quark mechanism for the production of a pair of $B_c$ mesons in the decay of the 
H boson is investigated. There are other production mechanisms, which are determined, for example, 
by the initial decay of the H boson into a pair of ZZ, WW and by other couplings. Our estimates 
of the contributions 
of such processes to the decay width show that they are two orders of magnitude smaller than 
the mechanism studied in our work.

The main parameters that give the theoretical error of the results obtained are the quark masses, 
the coupling constant $\alpha_s $ and the discarded corrections of order $O({\bf p}^4)$, 
$ O({\bf q}^4)$. In the case of the Higgs boson decay amplitudes
we use for the strong coupling constant $\alpha_{c}$
the two-loop approximation from \eqref{26} where the renormalization scale 
$\mu=\frac{m_1}{m_1+m_2}M_H$ and for $\alpha_{b}$ $\mu=\frac{m_2}{m_1+m_2}M_H$.
The correction of order $O({\bf p}^4)$ gives the most significant uncertainty in the 
theoretical value of the obtained decay widths \eqref{gapp}, \eqref{gavv}. Our total theoretical 
error of calculations is about 20 percent. 

As known the one loop QCD corrections can essentially contribute to the paired quarkonia 
production. Therefore  these corrections are also taken into account in the current study. 
We have found that the one loop contribution increases the width values by 1.3 -- 2.3 times.
In our model the relativistic corrections and one-loop effects act in different directions changing 
the Higgs boson decay widths obtained in the nonrelativistic  approximation. 
As a result, it turns out that at the $\mu=m_1 M_H/(m_1+m_2)$ 
energy scale, the total decay widths into the pair of pseudoscalar and vector $B_c$ mesons 
are the following: $\Gamma_{PP}^{tot}=1.0\cdot 10^{-12}~\text{GeV}$ and
$\Gamma_{VV}^{tot}=0.3\cdot 10^{-12}~\text{GeV}$.

\acknowledgments
We are grateful to A.~L.~Kataev for useful remarks.
The work of I.~N.~Belov and F.~A.~Martynenko is supported by the Foundation for the Advancement of Theoretical Physics and Mathematics "BASIS" (grants No. 20-2-2-2-1 and No. 19-1-5-67-1). The work of A.~V.~Berezhnoy and A.~K.~Likhoded is partially supported by RFBR (grant No.  20-02-00154 A). 

\appendix
\section{General structure of paired $B_c$ meson production relativistic amplitudes in the leading order in Higgs boson decay}
\bga
\label{eq:A1}
\mathcal M=\frac{4\pi}{3}M_{B_c}\times\\
\int\!\frac{d\mathbf{p}}{(2\pi)^3}\int\!\frac{d\mathbf{q}}{(2\pi)^3}
\frac{\Psi^0_{B_c}({\bf p})}
{\sqrt{\frac{\epsilon_1(p)}{m_1}\frac{(\epsilon_1(p)+m_1)}{2m_1}\frac{\epsilon_2(p)}{m_2}
\frac{(\epsilon_2(p)+m_2)}{2m_2}}}
\frac{\Psi^0_{B_c}({\bf q})}
{\sqrt{\frac{\epsilon_1(q)}{m_1}\frac{(\epsilon_1(q)+m_1)}{2m_1}\frac{\epsilon_2(q)}{m_2}
\frac{(\epsilon_2(q)+m_2)}{2m_2}}}\times \\
\mathrm{Tr}\bigl\{\mathcal T_{12}+\mathcal T_{34}\bigr\},
\ega
\bga
\mathcal T_{12}=\Gamma_c\alpha_{b}\Bigl[\frac{\hat v_1-1}2+\hat v_1\frac{\mathbf p^2}
{2m_2(\epsilon_2(p)+m_2)}-\frac{\hat p}{2m_2}\Bigr]\Sigma^{(1)}_{P,V}(1+\hat v_1)\times\\
\Bigl[\frac{\hat v_1+1}2+\hat v_1\frac{\mathbf p^2}{2m_1(\epsilon_1(p)+m_1)}+\frac{\hat p}{2m_1}
\Bigr]\left[\frac{\hat p_1-\hat r+m_1}{(r-p_1)^2-m_1^2}\,\gamma_\mu+\gamma_\mu\,\frac{\hat r-\hat q_1+m_1}{(r-q_1)^2-m_1^2}\right]D^{\mu\nu}(k_2) \\
\Bigl[\frac{\hat v_2-1}2+\hat v_2\frac{\mathbf q^2}{2m_1(\epsilon_1(q)+m_1)}+\frac{\hat q}{2m_1}
\Bigr]\Sigma^{(2)}_{P,V}(1+\hat v_2)
\Bigl[\frac{\hat v_2+1}2+\hat v_2\frac{\mathbf q^2}{2m_2(\epsilon_2(q)+m_2)}-\frac{\hat q}{2m_2}\Bigr]\gamma_\nu,
\ega
\bga
\mathcal T_{34}=\Gamma_b\alpha_{c}\Bigl[\frac{\hat v_1-1}2+\hat v_1
\frac{\mathbf p^2}{2m_1(\epsilon_1(p)+m_1)}+\frac{\hat p}{2m_1}\Bigr]\Sigma^{(1)}_{P,V}(1+\hat v_1)\\
\Bigl[\frac{\hat v_1+1}2+\hat v_1\frac{\mathbf p^2}{2m_2(\epsilon_2(p)+m_2)}-\frac{\hat p}{2m_2}\Bigr]
\left[
\frac{\hat p_2-\hat r+m_2}{(r-p_2)^2-m_2^2}\,\gamma_\mu+
\gamma_\mu\frac{\hat r-\hat q_2+m_2}{(r-q_2)^2-m_2^2}\,
\right]
D^{\mu\nu}(k_1) \\
\Bigl[\frac{\hat v_2-1}2+\hat v_2\frac{\mathbf q^2}{2m_2(\epsilon_2(q)+m_2)}-
\frac{\hat q}{2m_2}\Bigr]\Sigma^{(2)}_{P,V}(1+\hat v_2)
\Bigl[\frac{\hat v_2+1}2+\hat v_2\frac{\mathbf q^2}{2m_1(\epsilon_1(q)+m_1)}+
\frac{\hat q}{2m_1}\Bigr]\gamma_\nu,
\ega
where $\Sigma^{(1),(2)}_{P,V}$ is equal to $\gamma_5$ for pseudoscalar $B_c$ meson and
$\hat\varepsilon_{\mathcal{V}}$ for vector $B_c$ meson.
$\Gamma_c=m_c(\sqrt{2} G_F)^{1/2}$,
$\Gamma_b=m_b(\sqrt{2} G_F)^{1/2}$.
The trace calculation in \eqref{eq:A1}
leads to amplitudes ${\cal M}_{PP}$ and ${\cal M}_{VV}$ 
presented in~\eqref{eq:amp11}-\eqref{eq:amp22}.

\end{document}